# Addressing Components' Evolvement and Execution Behavior to Measure Component-Based Software Reliability

Wen-Li Wang
*School of Engineering*
*Penn State University, Behrend College Erie*
*wxw18@psu.edu*

Mei-Huei Tang
*Computer and Information Science Dept.*
*Gannon University*
*tang002@gannon.edu*

## Abstract

*Software reliability is an important quality attribute, often evaluated as either a function of time or of system structures. The goal of this study is to have this metric cover both for component-based software, because its reliability strongly depends on the quality of constituent components and their interactions. To achieve this, we apply a convolution modeling approach, based on components' execution behavior, to integrate their individual reliability evolvement and simultaneously address failure fixes in the time domain. Modeling at the component level can be more economical to accommodate software evolution, because the reliability metric can be evaluated by reusing the quality measures of unaffected components and adapting only to the affected ones to save cost. The adaptation capability also supports the incremental software development processes that constantly add in new components over time. Experiments were conducted to discuss the usefulness of this approach.*

## 1. Introduction

Component-based software has become popular for its ability to reuse and extend. Reusability improves productivity to the development of new systems while extendibility eases the constant evolvement of the existing systems. Component-based software development often faces a challenge of selecting reusable components and improving out-of-date ones. Even though frequent upgrades or updates can affect components' behavior and change their failure intensity, some components may not be changed so that their historical failure data can still be used to support quality measurement for the system. In order to economically re-evaluate reliability, it is better to combine those upgraded with the unchanged ones. The approach should also consider software repair to failures in order to accurately predict software reliability.

Software reliability models commonly adopt a black-box or a white-box approach. The black-box type models in [4,9,12,13] have the ability to predict reliability growths over time with the consideration of software repair. However, they treat a system as a whole without addressing the internal components and structures of software. The white-box type models [2,3,10,11,15] on the contrary tackle internal structures but are not time domain models to address software repair over time. For component-based software, the constant component evolvements as well as structural changes look for the strengths of both types.

In [6,18], component-based time domain models are introduced. They estimate reliability from the utilization of components and compute the system failure intensity as the sum of component failure intensities. Unlike path-based models [2,14,16], software architecture is only implicitly addressed regardless of the execution behavior. In [5], an ENHPP approach estimates components' reliabilities following their individual expected execution time. However, ENHPP does not address repair and predict reliability up to infinite time.

To address the above, we apply *convolutions* [1] to capture the benefits of both time domain and path-based models. The objective is to take into account components' progression and execution behavior while maintaining in the time domain to consider software repair. Software components are modeled individually based on their own failure time data to estimate component reliabilities. Component interactions are then employed to integrate the reliabilities of components running on paths through convolutions. Paths' last failure occurrence time will be used to account for repair following the Markov properties. Finally, the probability of traversing each path is addressed to compute software reliability. The popular non-homogeneous Poisson process (NHPP) type model [4,12] will be chosen for discussion.

In this paper, Section 2 describes our convolution approach. Section 3 applies it to the popular NHPP time domain model. Section 4 computes reliability and concerns software evolution. Section 5 discusses our model advantages, and related work is listed in Section 6. Section 7 contains some concluding remarks.

## 2. The Convolution Approach

Convolution [1] is our solution to bridge the gap between the white-box and the black-box approaches. This mathematical operator has been widely applied to relate system inputs and outputs, and to perform correlation. In [17], we have applied convolutions to estimate software's moving average (MA) [7] reliability growth in the discrete time domain. The result can help evaluate cost performance and support decision makings on future software improvements but consider no fault fixes. This section extends the scope to the continuous time domain in order for black-box type models to account for software repair, system strucutres, and components' behavior and evolvement.

An application of convolution is to compute the coefficients of the product of two polynomials, because it can relieve the computation burden for large vectors. Given vectors $f = [a\ b\ c]$ and $g = [d\ e]$ to represent two polynomials $ax^2 + bx + c$ and $dx + e$, the convolution $f * g$ in Eq. (1) yields $[ad\ bd+ae\ cd+be\ ce]$, where $n$ is for the $n$th coefficient result.

$$(f * g)(n) = \sum_{k=-\infty}^{\infty} f_k\, g_{n-k} \qquad (1)$$

The integration of components' evolvement follows the same idea, because components can be individually improved. Let $f$ and $g$ now be the reliability growths of components $c_f$ and $c_g$ as shown in Fig. 1. The MA reliability is equal to $[ad/1\ (bd+ae)/2\ (cd+be)/2\ ce/1]$, where the division averages the number of operands being summed. In [17], we formulated the MA reliability for discrete time intervals by Eq. (1) to be

$$MA(n) = (f * g)(n) ./ (\overline{f} * \overline{g})(n) \qquad (2)$$

where "./" represents component-wise vector division and $\overline{f} = [1\ 1\ 1]$ and $\overline{g} = [1\ 1]$. $\overline{f}$ and $\overline{g}$ are for the division purpose and have same lengths as $f$ and $g$.

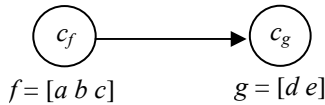

$f = [a\ b\ c]$        $g = [d\ e]$

**Fig. 1: A sequential system with two components**

In contrast to Eq. (1), for continuous functions convolution is the integral of the product of the two functions after one is reversed and shifted.

$$(f * g)(\tau) = \int f(t) g(\tau-t) dt \qquad (3)$$

Regarding functions $\overline{f}$ and $\overline{g}$, their values are always equal to 1. In the time domain, we have

$$(\overline{f} * \overline{g})(\tau) = \int 1\, dt \qquad (4)$$

For $f^{(1)}, f^{(2)}, \ldots, f^{(k)}$ to represent the reliabilities of $k$ sequential running components $c_1, c_2, \ldots, c_k$, the integrated MA reliability can be computed as

$$MA(\tau) = (f^{(1)} * f^{(2)} * \ldots * f^{(k)})(\tau) ./ \underbrace{\int \cdots \int}_{k} \underbrace{1\, d\tau \cdots d\tau}_{k} \qquad (5)$$

Eq. (5) can now tackle an execution path but has not yet addressed software with repair, whose formulation is subject to the adopted models. The next section will choose the popular NHPP model for application.

## 3. Application of MA to NHPP

NHPP [4,12] is one of the most popular software reliability growth models. It is a time domain finite failures Poisson type model that considers software with repair. The non-homogeneity of NHPP enables the consideration of variable failure intensity over time, an intrinsic property for accounting for repair. However, its black box feature limits the consideration of internal system structures. Our convolution modeling approach for computing MA intends to break this barrier to facilitate software evolution and be precise.

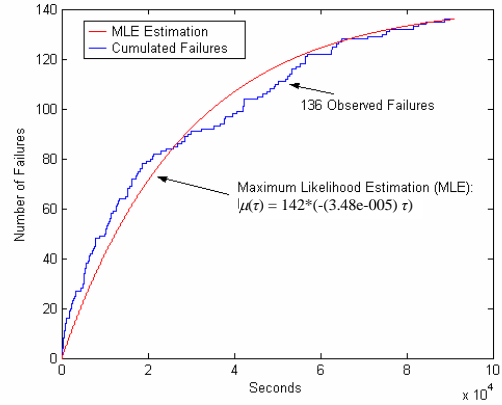

Fig. 2: The estimated $\mu(\tau)$ using MLE

The following shows our application of MA to NHPP models. A system was identified with 136 failures. In Fig. 2, the expected number of failures $\mu(\tau)$ experienced over execution time $\tau$ is estimated using the maximum likelihood estimation (MLE) [12] to be

142*(1-exp(-(3.48e-5) $\tau$)). The system's end of test time is at 91208 CPU second and its last failure occurrence time is at 88682 CPU second.

This system internally has two components $c_1$ and $c_2$ running sequentially. By MLE, component $c_1$ has $\mu_1(\tau)$ equal to 69*(1-exp(-(4.3553e-5) $\tau$)) while $c_2$ has $\mu_2(\tau)$ equal to 74*(1-exp(-(2.7482e-5) $\tau$)). The MA reliability computed from our convolution approach will be used to integrate components' $\mu_1(\tau)$ and $\mu_2(\tau)$ to estimate the system's $\mu(\tau)$. This is different from the additive models [18]. Our paradigm does not compute the system failure intensity as the sum of component failure intensities, $\mu(\tau) = \mu_1(\tau) + \mu_2(\tau)$. Instead, only the components' scale parameters 69 and 74 are summed to represent the system's scale parameter $v_0 = 143$, but the interactions between components are convolved and averaged to yield a macroscopic MA reliability. This helps account for all kinds of components' execution time combinations, i.e., being able to consider their dynamic execution behavior. Therefore, if a failure occurred at the *n*th second, our approach addresses not just any single component that ran and failed at that time, but also multi-components that ran a total of *n* seconds and failed.

Based on the above, the system $\mu(\tau)$ is derived from $\mu_1(\tau)$ and $\mu_2(\tau)$ using MA($\tau$) as follows. By NHPP, $c_1$'s unconditional reliability is $f = \exp(-(4.3553e-5) \tau)$ and $c_2$'s is $g = \exp(-(2.7482e-5) \tau)$. By Eq. (5), MA($\tau$) is computed as $(f - g) / [(-2.7482e-5+4.3553e-5) \tau]$. With both $v_0$ and MA($\tau$), $\mu(\tau)$ is computed as

$$\mu(\tau) = v_0 (1 - MA(\tau)) \quad (6)$$

Software reliability $R(\tau_i' \mid \tau_{i-1})$ with the consideration of repair is a conditional reliability of time $\tau_i'$ under the last failure time at $\tau_{i-1}$. It is formulated in [12] as

$$R(\tau_i' \mid \tau_{i-1}) = \exp(-(\mu(\tau_{i-1}+\tau_i') - \mu(\tau_{i-1}))) \quad (7)$$

Fig. 3 shows the $\mu(\tau)$ plot of those three different approaches for comparison. The additive and MA approaches both have a close result to the NHPP. The MA curve is slightly above the NHPP, while the additive curve is a little below it. There is no clear evidence about which one gets a close reliability match with the NHPP. With the last failure occurrence time at 88682 second, Figure 4 shows the conditional reliability plots by Eq. (7). It can be seen that the MA and the NHPP results have a very close overlap, while the additive result does not. This is because the results of MA and NHPP are with the consideration of components' dynamic interactions. The NHPP addresses the known interactions through the external test cases, while our MA takes advantage of convolutions to mimic all sorts of possible internal execution behavior between component interactions.

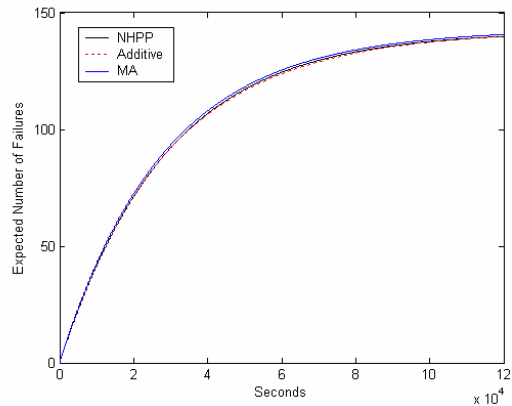

Fig. 3: The estimated $\mu(\tau)$ of the three approaches

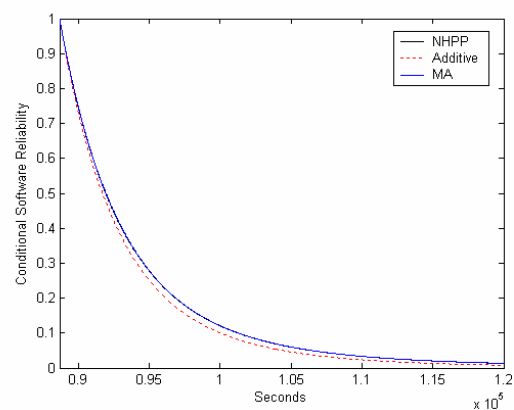

Fig. 4: Comparison of $R(\tau_i' \mid \tau_{i-1})$ results

## 4. Software Reliability Modeling and Software Evolution

Section 3 has demonstrated the applicability of MA, computed based on our convolution approach. However, the modeled system was rather simple that had only one execution path and two components. For software, it is likely to have many execution paths and each can have a different set of components running. Furthermore, a component may also appear in multiple execution paths. Therefore, software evolution like upgrades or updates to a component will possibly affect several paths, but not definitely all of them.

The MA in Eq. (5) is already capable of handling an execution path with more components. For *n* different components on a path, there will be *n*-1 convolutions performed on the numerator and another *n*-1 on the denominator. Repeated component occurrences on a path can be neglected, because the computations will not change the MA result. For a system with multiple execution paths, software reliability is computed as the sum of the products of the conditional reliability of each path and its corresponding traversal probability. The traversal probability $P_\varepsilon$ for a path $\varepsilon$ can be yielded from test or an operational profile. Let $R_\varepsilon$ represent the reliability of path $\varepsilon$. For $m$ paths, the conditional software reliability $R$ is formulated as

$$R(\tau_i' \mid \tau_{i-1}) = \sum_{\varepsilon=1}^{m} R_\varepsilon(\tau_i'+\tau_{i-1}-\tau_\varepsilon \mid \tau_\varepsilon) \cdot P_\varepsilon \qquad (8)$$

Because the last failure occurrence time $\tau_{i-1}$ of the system can be different from the last failure occurrence time $\tau_\varepsilon$ of a path, Eq. (8) forces all paths to be modeled at the same time $\tau_i'+\tau_{i-1}$.

Our paradigm addresses system structures through paths. Since software evolution may not affect all paths, there is no need for our approach to re-measure the unaffected ones, thus saving time and effort. Even for an affected path, only the affected components or configurations need to be tackled. In contrast, the black-box characteristic of NHPP will need to re-test the whole system for a component replacement, because the scale of change is unlike a repair to be small. Consequently, the past failure data are no longer trustworthy and can result in inaccurate measurement.

For example, component $c_2$ in Section 3 is replaced with another one with a new $\mu_2(\tau)$. The component interactions remain the same, i.e., $c_1$ connects to the new component. Our convolution approach can quickly re-compute MA for the new reliability, while NHPP needs to re-test the whole system in order to get a global $\mu(\tau)$. This can be difficult and expensive when software continues to evolve.

## 5. Discussions

The two components in Section 3 happened to have very close failure intensity. They also have the same end of test time as the system model. As a consequence, the reliability trend matched tightly with that of NHPP. However, in practice these are often not the case. The failure intensity of each component can differ and the test time to individual components may also vary. Our convolution approach provides an advantage of addressing component interactions that allows components to differ in test time and failure intensity. This is however a challenge for NHPP or additive models.

### 5.1 Different failure intensity of components

Components can differ enormously in their number of failures over time, because some are likely to be more fault prone than others. The black-box characteristic of NHPP treats a system as a whole without concerning its constituent components and their differences. For our convolution approach, software reliability remains a function of time to address repair. More importantly, the model can consider components' interactions and their execution time combinations.

Take Figure 1 for example. Eq. (2) integrates vectors $f$ and $g$ of components $c_f$ and $c_g$ into a four elements MA vector by convolutions. The MA now covers the time span of the *sum* of execution times of the two components. Assume the discrete sampling is done in every second. The three values of $f$ are sampled at seconds 0, 1, and 2 while the two of $g$ are done at 0 and 1. An integrated system of the two components will have MA values at seconds 0, 1, 2, and 3. The 3 is the summation result of 2 seconds of execution of $c_f$ and 1 second of execution of $c_g$. Table 1 lists the execution time combinations.

|   | $c_f$ | $c_g$ |
|---|---|---|
| 0 | 0 | 0 |
| 1 | 0 | 1 |
|   | 1 | 0 |
| 2 | 1 | 1 |
|   | 2 | 0 |
| 3 | 2 | 1 |

Table 1: Execution time combinations of $c_f$ and $c_g$

Similarly, continuous time convolutions of Eq. (5) are applicable to compute MA for continuous functions. Assume a system has $n$ sequential components and each component's $f^{(i)}$ reliability function is yielded based on its failure time data up to $s_i$ seconds. The execution time combinations will range from 0 to $\sum s_i$ seconds for the MA reliability function. The expected number of failures $\mu(\tau)$ is also yielded based on the time span of MA($\tau$), as shown in Eq. (6), to adhere closely to the execution behavior. Fig. 5 shows an experimental result that our model can capture the impact caused by uneven failure intensity of components.

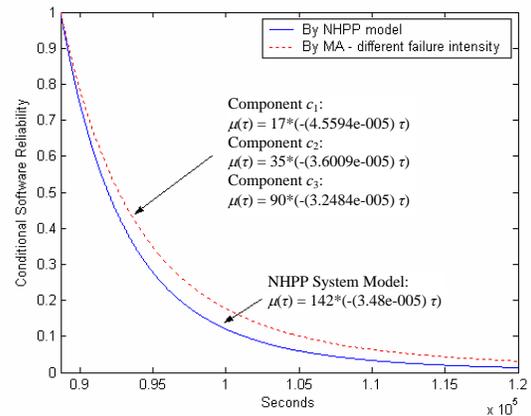

Fig. 5: Diverse failure intensity of the components

For additive models, the $\mu(\tau)$ covers only up to the observed longest failure time of the components on an execution path. Components' execution time combinations are not addressed to explicitly tackle component interactions and runtime behavior. In this case, our convolution approach offers an advantage.

### 5.2 Varied end of test time to components

The dissimilarity among components is not limited to their inherent difference in failure intensity. Components can also result in a different end of test time. For instance, we would prefer to spend less energy to test a non-critical component than a critical one, and complete test early for a simple component than a complex one. It is not cost effective to treat all components as alike by allocating them the same test time and effort. This will result in components being under test or over test. An appropriate way is to assign more resources to those that are more error prone or with high failure intensity.

Based on the previous experimental setup in Fig. 5, another experiment was conducted to perform different test time duration to the components. The first two components now have a shorter test time that ends at 10240 and 60000 CPU seconds. The third component remains the same to have 91208 CPU seconds test time. Fig. 6 shows the new reliability plot. It is clear to see that there is a significant enhancement. The reason is because components with a shorter test time are mainly those who have become hard to come across another failure. Since the future chance to encounter a failure in them is relatively less, the conditional software reliability is higher. The black box NHPP cannot address this issue, so its curve stays the same. In regard to additive models, it is impractical to sum the failure intensities of components to get the system's failure intensity from inconsistent test time durations.

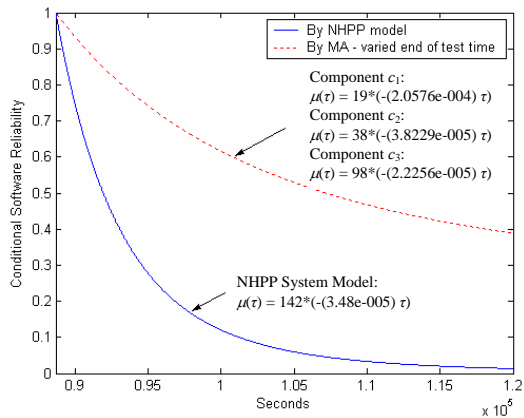

Fig. 6: A varied end of test time to components

## 6. Related Work

The introduction in Section 1 has discussed a number of related studies to our work. For a similar technique being used, Laprie and Kanoun's reliability model [8] is the closest approach to ours. It is also a time domain model and utilizes the convolution operator to integrate software components. Instead of computing the MA, components' failure probability density functions are convolved to gain a global density function to compute software reliability. However, two potential problems can hinder the use of this approach. First, the integration can only be performed after all components have failure data. Otherwise, a failure free component will make the convolution result being all 0's, giving a wrong impression that the system is totally reliable. Secondly, the convolution computes the product of components' failure probability density functions, i.e., the global density function considers only the probability that the interacting components all fail. Its use can result in reliability over estimate.

## 7. Conclusions

We have developed and applied the convolution approach to a continuous time reliability model to simultaneously address repair, software evolution, and system structures for component-based software. Convolutions help bridge the gap between the black-box and white-box approaches, enabling execution behavior to be modeled for accuracy, component upgrades and updates to be considered for evolution, and continuous time to be maintained for constant repair. The convolution approach integrates component reliabilities to yield an MA result, which is then used to estimate the expected number of failures and conditional reliability for an execution path. Finally, software reliability is computed by further considering the traversal probability of each path.

The integration capability allows components to be modeled individually to match with their best fit mean value and failure intensity functions. In addition, components can undergo different test time and their possible execution behavior covers all sorts of execution time combinations to be thorough. The model's easy adaptation to software evolution without a need of retesting the complete system saves time and effort. Experiments were conducted to compare our MA result with that of the NHPP model. One demonstrates the ability of our paradigm to take into account components' diverse failure intensities, and the other further shows the impact of having different test time to components on software reliability. The limitations of the additive models are also discussed.

In summary, the advantage of our approach has many folds. First, it is useful by considering more factors. Second, the characteristics of individual components can be addressed to best describe the system. Third, the component-level modeling can save cost on software evolution by reusing unaffected measures. Fourth, it facilitates incremental software processes to be exercised for adding new components. In the future, we will study its application to other types of models and explore the feasibility of modeling software with heterogeneous architecture.